\documentclass[prl,twocolumn,showpacs]{revtex4}
\usepackage[latin1]{inputenc} 
\usepackage[T1]{fontenc} 
\usepackage{textcomp,lmodern} 
\usepackage{amssymb, amsmath} 
\usepackage{ntheorem}
\DeclareMathOperator{\Spur}{Tr} %
\newcommand{\ens}[0]{\ensuremath} 
\newcommand{\anfEngl}[1]{``#1''} 
\newcommand{\etal}[0]{et al.} 
\newcommand{\ket}[1]{\ens{|#1\rangle}} 
\newcommand{\bra}[1]{\ens{\langle#1|}} 
\newcommand{\x}[0]{\ens{\otimes}} 
\newcommand{\Mge}[2]{\ens{\left\lbrace #1|\,#2 \right\rbrace}} 
\newcommand{\Mg}[1]{\ens{\left\lbrace #1 \right\rbrace}} 
\newcommand{\MgN}[1]{\ens{\Mg{0,\dots,#1}}} 
\newcommand{\MgE}[1]{\ens{\Mg{1,\dots,#1}}} 
\newcommand{\betrag}[1]{\ens{|#1|}} 
\newcommand{\pr}[1]{\ens{\ket{#1}\bra{#1}}} 
\newcommand{\nGr}[0]{\ens{{n \rightarrow \infty}}}
\newcommand{\cH}[0]{\ens{\mathcal{H}}}
\newcommand{\fU}[0]{\ens{\mathfrak{U}}}
\newcommand{\N}[0]{\ens{\mathbb{N}}} 

\newcommand{\C}[0]{\ens{\mathbb{C}}}
\newcommand{\Bnd}[0]{\ens{B_n^{(d)}}}
\newcommand{\iE}[0]{\ens{\mathrm{i}}} 
\newcommand{\Alm}[0]{\ens{(A_{lm})_{l,m = 0}^{d-1}}}
\newcommand{\Aplm}[0]{\ens{(A^\prime_{lm})_{l,m = 0}^{d-1}}}
\newcommand{\pers}[1]{#1} 
\newcommand{\aut}[2]{#1 #2} 
\newcommand{\autgr}[2]{#1 and #2} 
\newcommand{\zeit}[1]{#1} 
\newcommand{\Band}[1]{\textbf{#1}} 
\newcommand{\Seiten}[2]{\mbox{#1--#2}} 
\newcommand{\ITIT}[0]{IEEE Trans. Inf. Th.}
\newcommand{\JPA}[0]{J. Phys. A}
\newcommand{\PRA}[0]{Phys. Rev. A}
\newcommand{\PRL}[0]{Phys. Rev. Lett.}

\newtheorem*{Definition}{Definition} 
\newtheorem{Theorem}{Theorem}

\begin{document}
\title{Symmetric extendibility for qudits and tolerable error rates in quantum cryptography}
\author{Kedar S. Ranade}
\affiliation{Institut für Angewandte Physik, Technische Universität Darmstadt, 64289 Darmstadt, Germany}

\date{May 31, 2009}

\begin{abstract}
  Symmetric extendibility of quantum states has recently drawn attention in the context of quantum cryptography to judge
  whether quantum states shared between two distant parties can be purified by means of one-way error correction protocols.
  In this letter we study the symmetric extendibility in a specific class of two-qudit states, i.\,e. states composed
  of two $d$-level systems, in order to find upper bounds on tolerable error rates for a wide class of qudit-based
  quantum cryptographic protocols using two-way error correction. In important cases these bounds coincide with
  previously known lower bounds, thereby proving sharpness of these bounds in arbitrary finite-dimensional systems.
\end{abstract}

\maketitle
Quantum cryptography or quantum key distribution was introduced by the seminal work of Bennett and Brassard in 1984
\cite{BB84}. It was widely conjectured to provide means for unconditionally secure communication of
two distant parties, Alice and Bob, even in the presence of an adversary, Eve, who can be detected
by the quantum-mechanically-induced noise every attack produces. However, security proofs still left loopholes,
until Mayers \cite{Mayers} came up with a proof which provided unconditional security even for some finite amount
of noise. Since every real physical channel inevitably produces noise, however low it may be, this proof amounts
to the advent of practically useful quantum cryptography. Starting from this work, the security of quantum
cryptography  has been investigated in various scenarios and its security was proved therein.
Whereas most of the efforts have been performed with respect to two-dimensional systems (\emph{qubits}
with Hilbert space $\cH = \C^2$) as basic quantum systems, a large number of concepts and methods carry through for
arbitrary finite-dimensional systems (\emph{qudits} with Hilbert space $\cH = \C^d$).
Such general finite-dimensional systems will be the main focus of this letter. However, we do not consider
infinite-dimensional systems (continuous-variable quantum key distribution), since the protocols and scenarios
employed there differ considerably from the finite-dimensional case.
\par A particularly important task in quantum-cryptography-related research is to identify maximally tolerable error
rates, i.\,e. the error rates measurable by Alice and Bob, up to which a physical channel may be used for quantum
key distribution without compromising the unconditional security. For the BB84 protocol, this question can be
answered by considering an entanglement-based version \cite{Ekert} of that protocol with an appropriate choice of
entanglement purification. Under particular circumstances, this protocol can be reduced to yield security of the
original BB84 protocol itself by transforming the entanglement purification to some post-processing, namely error
correction and privacy amplification. One generically distiguishes two types of post-processing.
The first kind only uses one-way classical communication, i.\,e. in the course of post-processing
Alice may send classical information to Bob, but not vice versa; Shor and Preskill used such protocols to
show security of the BB84 protocol up to an error rate of 11.0\,\% \cite{ShorPreskill}.
The second kind of post-processing uses both one- and two-way classical communication \cite{GottesmanLo}, and
this can be used to derive tolerable error rates of 20\,\% for the BB84 protocol \cite{Chau02,Acin,RanadeAlber-1}.
(One should note, that by one-way protocols only, it is impossible to exceed a rate of about 15\,\%, since an
arbitrary qubit state can be duplicated, if this rate of error is allowed.) By now no protocol is known which
works above 20\,\% error rate, although it has been shown \cite{NikolopoulosKhaliqueAlber} that effective entanglement,
which is necessary for key generation, exists up to an error rate of 25\,\%. At present, there exists no purification
protocol which works between these two bounds, and it is unknown whether this gap can be closed at all.
\par A different approach to investigate this problem was recently undertaken by Myhr \etal{}
\cite{Myhr_ua,MyhrLuetkenhaus}, who posed the question, which condition a quantum state has to fulfil in order to be
correctable, using one-way classical communication only. They found that the concept of symmetric extendibility served
their purpose, i.\,e. that a state
which possesses a symmetric extension cannot be purified by any possible protocol with one-way communication only.
By working out a criterion for symmetric extendibility for two-qubit states they achieved to show that the
abovementioned bound of 20\,\% cannot be surpassed by standard means of two-way entanglement purification
\cite{GottesmanLo,Chau02,RanadeAlber-1,Acin}, which indicates the need of new two-way protocols. Finally,
we should note that all results we mentioned for the BB84 protocol carry over for
the six-state protocol \cite{Brusz}, if the rates of 20\,\% and 25\,\% are replaced by 27.6\,\% and 33.3\,\% \cite{Chau02,Acin,RanadeAlber-1,NikolopoulosKhaliqueAlber,Myhr_ua,MyhrLuetkenhaus}.
\par The aforementioned results show that the tolerable error rates of the most important qubit-based protocols
are known, at least in a reasonable scenario using the error correction and privacy amplification available
at present. One may ask whether similar results hold true for general finite-dimensional systems, in case we
use generalised BB84 and six-state schemes. While lower bounds for such cases are known
\cite{Chau05,NikolopoulosRanadeAlber,RanadeAlber-2,BaeAcin}, with the exception of disentanglement bounds \cite{NikolopoulosAlber}
we are not aware of a systematic study of upper bounds, which is the main purpose of this letter. By performing
an analysis similar to that of Myhr \etal, we will derive upper bounds on the tolerable error rates with two-way
communication. As Myhr et al. did for qubit-based protocols, we show that these upper bounds coincide with
the lower bounds already known \cite{Chau05,NikolopoulosRanadeAlber,RanadeAlber-2,BaeAcin}, thus proving sharpness of
these bounds. In the following we will state the details of the model and our proof.
\par Let us start by introducing the concept of symmetric extendibility and its relevance in quantum cryptography \cite{Myhr_ua,MyhrLuetkenhaus,Terhal_ua}.
\begin{Definition}[Symmetric extendibility]\hfill\\
  A state $\rho_{AB}$ on the tensor product $\cH_A \x \cH_B$ of two Hilbert spaces is said to be
  \emph{symmetrically extendible}, if there exists a tripartite state $\rho_{ABE}$ on $\cH_A \x \cH_B \x \cH_E$ with
  $\cH_E = \cH_B$, such that $\Spur_E \rho_{ABE} = \rho_{AB}$ (extendibility) and $\rho_{ABE} = \rho_{AEB}$ (symmetry)
  hold.
\end{Definition}
Obviously all separable states possess symmetric extensions, whilst no pure entangled state can be extended.
The general solution to the problem, whether a state is symmetrically extendible or not is unsolved, however, a criterion
for Bell-diagonal two-qubit states is known \cite{Myhr_ua} and, more generally, criteria for general two-qubit states
have been investigated \cite{MyhrLuetkenhaus}. The relevance of symmetric extendibility in quantum cryptography
arises from the following observation \cite{Myhr_ua}: If Alice and Bob share a symmetrically extendible state, we cannot
exclude the possibility that Eve holds the extension. If Alice then tries to use one-way communication to establish
a secret key with Bob, she will fail, since Eve could do precisely the same as Bob, and Bob and Eve are indistinguishable
to Alice.
\par We shall  now introduce a class of two-qudit states, which was shown to be relevant in quantum cryptography.
For states within this class we have derived a criterion for symmetric extendibility (Theorem \ref{SymExt}), which
enables us to obtain a simple sufficient condition for this property. Based upon this condition, we will
derive upper bounds for qudit-based quantum cryptographic protocols and compare them with the previously known results.
To this end, we will in the course of this letter focus on a particular two-way error correction step, the $\Bnd$ step
(see below), which is essentially the only genuine two-way post-processing method used in quantum cryptography. Applying
this step to a chosen initial state, we will check whether we can reach a state which does not possess a symmetric extension.
If this is not the case, we conclude that no protocol in the class considered can produce a secret key for a particular
given error rate. The bounds which we derive lie below the disentanglement threshold \cite{NikolopoulosAlber}.
Thus, in order to achieve this threshold---if it is possible at all---new two-way methods have to be invented.
\par To determine whether a general state is symmetrically extendible or not is a complicated task, even
for two-qubit states \cite{MyhrLuetkenhaus}. However, in the context of entanglement-based quantum cryptography
by using arguments of the Gottesman-Lo type \cite{GottesmanLo} we may concentrate on a subclass of all states,
namely the (generalised) Bell-diagonal states \cite{Chau02,Chau05,GottesmanLo,Acin, RanadeAlber-1,NikolopoulosRanadeAlber,RanadeAlber-2} on
$\cH = \C^d \x \C^d$ where $d$ is the dimension of a single quantum system, shared by Alice and Bob.
This is achieved by a fictive-measurement argument \cite{GottesmanLo}: One can perform a measurement in the so-called
Bell basis before starting the actual protocol, and such measurement does not have any measurable effect on the key.
The Bell basis consists of vectors (we denote $z := \exp[2\pi\iE/d]$, and $\ominus$ is to be taken modulo~$d$)
\begin{equation}
  \ket{\Psi_{lm}} := d^{-1/2} \sum\nolimits_{k = 0}^{d-1} z^{lk} \ket{k} \ket{k \ominus m}.
\end{equation}
for $l,\,m \in \MgN{d-1}$, and the Bell-diagonal states can be written in the form
\begin{equation}\label{Belldiag}
  \rho = \sum\nolimits_{l,m=0}^{d-1} A_{lm} \pr{\Psi_{lm}}.
\end{equation}
These states are completely determined by their coefficient matrix $\Alm$. Choosing the perfect state to be
$\pr{\Psi_{00}}$---Alice and Bob can locally measure in the standard basis and get one perfectly correlated key dit---
we can interpret nonzero $l$ and $m$ to be phase and dit errors, respectively.
\par In quantum cryptography we may assume that Alice and Bob share a large number of identical Bell-diagonal states,
i.\,e. $\rho^{\x N}$ for Bell-diagonal $\rho$ and $N \gg 1$, which have to be processed by using one-way and two-way communication to yield the state $\pr{\Psi_{00}}$ \cite{GottesmanLo}. Assuming that the state does not possess
a symmetric extension, we have to use two-way entanglement purification in order to break the symmetric
extension \cite{Myhr_ua,MyhrLuetkenhaus}. Apart from minor modifications, all known genuine two-way protocols
are variations of the $\Bnd$ step, so we may focus on that particular step \cite{RanadeAlber-2}:
Alice and Bob choose $n \in \N$ qudit pairs, all prepared in the state $\rho$ with coefficients $\Alm$. They then
locally apply generalised XOR operations $\ket{i}\ket{j} \mapsto \ket{i}\ket{i \ominus j}$ from the first to all
other pairs and afterwards measure the dit values of all pairs except
the first one. They compare the parities of all $n-1$ measurements by means of classical (two-way) communication
and keep the first pair, only if all parities coincide. If they keep the first pair, it will be described by
coefficients $\Aplm$, which are given by \cite{RanadeAlber-2}
\begin{equation}\label{BndEntw}
  A_{lm}^\prime = (dN)^{-1} \sum\nolimits_{i = 0}^{d-1} \, \Biggl[ \, z^{-il} \,
    \biggl(\sum\nolimits_{j = 0}^{d-1} z^{ij} A_{jm} \biggr)^n\Biggr],
\end{equation}
where $N := \sum_{m = 0}^{d-1} \bigl(\sum_{l = 0}^{d-1} A_{lm}\bigr)^n$. This state has
in general less dit errors but more phase errors than the original state by the properties of the generalised
XOR. Since we can assume an error rate below 50\,\% \cite{NikolopoulosAlber}, i.\,e.  $\sum_l A_{l0} > 1/2$, in the
limit of large $n \in \N$, the output state of the $\Bnd$ step lies in the neighbourhood of the separable state
defined by $A_{lm} = d^{-1}\delta_{m0}$.
\par We can even further reduce the set of states which have to be taken into consideration by enforcing their
invariance with respect to the Abelian unitary group $\fU_2 := \Mge{U \x U^*}{U \,\text{diagonal in
the standard basis}}$. This is due to the fact that the \Bnd step eliminates all kinds of error affected by this symmetrisation, so that they do not enter in the one-way correction part \cite{RanadeAlber-2}. Such $\fU_2$-invariant
Bell-diagonal states are characterised by the additional property that $A_{lm} = d^{-1}A_{*m}$ holds with
$A_{*m} := \sum_l A_{lm}$ for all $m \neq 0$.
\par One can now try to explicitly construct a symmetric extension of the given state as a $d^3 \times d^3$ matrix.
The $\fU_2$-invariance implies that a symmetric extension, if it exists at all, can be chosen to be
invariant with respect to $\fU_3 := \Mge{U \x U^* \x U^*}{U \,\text{diagonal in the standard basis}}$, and this leads to
a block matrix structure of the extension. By some further processing, one can work out a criterion for
symmetric extendibility for $\fU_2$-invariant Bell-diagonal states. Defining $\tilde{A}_{ip} := \sum_{l} A_{l0} z^{l(i-p)}$,
this criterion reads as follows \cite[p. 91]{TeilA,Ranade}.
\begin{Theorem}[Symmetric extendibility]\label{SymExt}\hfill\\
  A $\fU_2$-invariant Bell-diagonal state is symmetrically exten\-dible, if and only if
  $(d^{-1}\tilde{A}_{ip})_{ip} \in \C^{d \times d}$ can be decomposed into the sum of $d$ matrices
  $B_0,\,B_1,\,\dots,\,B_{d-1}$, such that every matrix $B_k = (a^{(k)}_{ip})_{ip}$
  is positive and $a^{(k)}_{ii} \leq d^{-1}A_{*,i \ominus k}$ for all $i,\,k \in \MgN{d-1}$.
\end{Theorem}
Although this condition is in general difficult to check, it turns out that it is sufficient for calculating
upper bounds on tolerable error rates in quantum cryptography. To this end, we have to find an explicit
solution to Theorem~\ref{SymExt} nearby the state with coefficients $A_{lm} = d^{-1}\delta_{m0}$ as mentioned
before.
\par For such a state $A_{*0} \approx 1$ and $A_{*m} \approx 0$ for $m \neq 0$ hold. Comparing this with
Theorem \ref{SymExt}, we conclude that in every matrix $B_k$ one of the diagonal elements can be large,
while all others have to be small. Thus we are tempted to choose the matrix e.\,g. $B_0$ to have the form
\begin{equation}\label{Matrix}
  \begin{pmatrix}
    \alpha     & \eta_1^* & \eta_2^* & \hdots & \eta_{d-1}^* \\
    \eta_1     & \beta_1  & 0        & \hdots & 0 \\
    \eta_2     & 0        & \beta_2  & \ddots & \vdots \\
    \vdots     & \vdots   & \ddots   & \ddots & 0 \\
    \eta_{d-1} & 0        & \hdots   & 0      & \beta_{d-1}
  \end{pmatrix} \in \C^{d \times d},
\end{equation}
where $\alpha = a_{00}^{(0)}$ may be large and the $\beta_i = a_{ii}^{(0)}$ have to be small. For the other
$B_k$ we assume a similar structure, but $\alpha$ and the $\eta_i$ lie on the $(k-1)$-th row and column.
Since the matrix has to be positive, and thus every $2 \times 2$ principal minor must be non-negative, we may
say that the elements set to zero have to be quadratically small in $\beta_i^{1/2}$, whilst the $\eta_i$ are only
linearly small.
\par The point of enforcing this matrix structure is that it is positive, if and only if (a) all diagonal
elements are non-negative and (b) its determinant is non-negative. This follows by direct application
of the well-known Hurwitz-Sylvester criterion for positive semidefiniteness, which states that a Hermitian matrix
is positive semidefinite, if and only if all principal minors are non-negative; see e.\,g. \cite[p. 282]{Gantmacher}.
If all $\beta_k$ are strictly positive, we can write the determinant of the matrix in (\ref{Matrix}) as
\begin{equation}
  \left(\prod\nolimits_{l = 0}^{d-1} \beta_l\right)
           \times \left(\alpha - \sum\nolimits_{k = 1}^{d-1} \frac{\betrag{\eta_k}^2}{\beta_k} \right),
\end{equation}
and we may focus on the right factor of that product.
\par We now want to construct a matrix $B_0$ of the form mentioned in eq. (\ref{Matrix}). The other matrices
$B_k$ shall have the same entries, but the row and column structure is permuted according to the conditions of
Theorem \ref{SymExt}. Note that the exceptional element $\alpha$ for these $B_k$ wanders along the diagonal with
increasing $k = 0,\,1,\,\dots,\,d-1$, but that does
not alter positivity. It seems appropriate to set the small values $\beta_m$ to their maximally possible value
$\beta_m := d^{-1}A_{*m}$ according to Theorem \ref{SymExt}, so normalisation enforces $\alpha := (2A_{*0}-1)/d$.
Since all diagonal elements are non-negative, we find
\begin{equation}\label{ineq}
  \sum\nolimits_{m = 1}^{d-1} \frac{\betrag{\eta_m}^2}{d^{-1}A_{*m}} \leq \frac{2A_{*0}-1}{d} = \frac{1 - 2(1-A_{*0})}{d}
\end{equation}
to be necessary and sufficient for positivity. (Given the case that some $A_{*m} = 0$, this is to be interpreted that
the coefficient $\eta_m$ must vanish.)
For a state to be symmetrically extendible, $\eta_m + \eta_{d \ominus m}^* = d^{-1}\tilde{A}_{m0}$
must hold in addition for all $m \in \MgE{d-1}$. Since in the inequality (\ref{ineq}), there appear only absolute
values, we may
set the phases of $\eta_m$ and $\eta_{d \ominus m}^*$ to be equal, for choosing small $\betrag{\eta_i}$ does not
harm positivity. We thus remain with the condition $\betrag{\eta_m} + \betrag{\eta_{d \ominus m}} = d^{-1} \betrag{\tilde{A}_{m0}}$. Setting $\betrag{\eta_m} =: \chi_m d^{-1}\betrag{\tilde{A}_{m0}}$, we can
rewrite the inequality as
\begin{equation}
   \sum\nolimits_{m = 1}^{d-1} \left[\chi_m^2 \frac{\betrag{\tilde{A}_{m0}}^2}{A_{*m}} + 2 A_{*m}\right] \leq 1,
\end{equation}
which is to be fulfilled under the additional constraint $\chi_m + \chi_{d \ominus m} = 1$.
\par Applying a $\Bnd$ step to the state, equation (\ref{BndEntw}) implies $A_{*m}^\prime = A_{*m}^n/N$
and $\tilde{A}_{m0}^\prime = \tilde{A}_{m0}^n / N$ with $N = \sum_m A_{*m}^n$, and the condition then reads
\begin{equation}
  \sum\nolimits_{m = 1}^{d-1} \left[\chi_m^2 \frac{1}{N} \left(\frac{\betrag{\tilde{A}_{m0}}^2}{A_{*m}}\right)^n
    + 2 \frac{A_{*m}^n}{N}\right] \leq 1.
\end{equation}
We shall ignore the second term in the limit $\nGr$, since it converges to zero. For the first term we can use the inequality
$N \leq A_{*0}^n$, which gets tight for $\nGr$. The term in question then is $\betrag{\tilde{A}_{m0}}^2/(A_{*0}A_{*m})$;
if it is less than $1$, it will disappear for $\nGr$, otherwise we have to suppress it by setting $\chi_m := 0$.
This will not be possible, if both $\betrag{\tilde{A}_{m0}}^2/(A_{*0}A_{*m})$ and
$\betrag{\tilde{A}_{m0}}^2/(A_{*0}A_{*,d \ominus m})$ are greater than $1$, which shows the following theorem.
\begin{Theorem}[Non-correctability]\hfill\\
  After the application of a $\Bnd$ step for sufficiently large $n$ to a $\fU_2$-invariant Bell-diagonal state with
  $A_{*0} > 1/2$, the output state is symmetrically extendible, if there hold the inequalities
  $\betrag{\tilde{A}_{m0}}^2 < A_{*0} \cdot \max\Mg{A_{*m},\,A_{*,d \ominus m}}$ for all $m \in \MgE{d-1}$.
\end{Theorem}
In the apparently-isotropic case, that is the case where we enforce $A_{*m} = (1-A_{*0})/(d-1)$ for all $m \neq 0$,
the condition reads
\begin{equation}\label{RatenKrit}
  \left(\max\nolimits_{m = 1}^{d-1} \betrag{\tilde{A}_{m0}}\right)^2 < A_{*0} \cdot \frac{1-A_{*0}}{d-1}.
\end{equation}
If we replace \anfEngl{<} by \anfEngl{>}, this is our sufficient condition for correctability \cite{RanadeAlber-2},
so that these two results are complementary (apart from the case of equality).
\par A particularly simple and instructive case is the generalised-isotropic case \cite{NikolopoulosRanadeAlber,RanadeAlber-2}.
In the $\fU_2$-invariant case, we have two non-negative real parameters $\alpha$ and $\beta$ which
fulfil $\alpha + (d-1)\beta \leq 1$, and the coefficients of the state are given by
\begin{equation}
  A_{lm} = \begin{cases}
             \alpha, & \text{if $l = m = 0$,}\\
             \beta, & \text{if $l \neq m = 0$,}\\
             \frac{1-\alpha-(d-1)\beta}{d(d-1)} & \text{else.}
           \end{cases}
\end{equation}
Denoting $x := \alpha + (d-1)\beta$, we thus compute
\begin{equation}
  \tilde{A}_{ip} = \begin{cases}
                     x,     &\text{if $i = p$,}\\
                     \alpha - \beta, &\text{else}
                   \end{cases},\,
    A_{*m} = \begin{cases}
                     x,               &\text{if $m = 0$,}\\
                     \frac{1-x}{d-1}, &\text{else;}
             \end{cases}
\end{equation}
inequality (\ref{RatenKrit}) now reads $(\alpha-\beta)^2(d-1) < x \cdot (1-x)$ and can be rewritten as
\begin{equation}
  \alpha^2 + (d-1) \beta^2 - \frac{\alpha + (d-1)\beta}{d} < 0.
\end{equation}
This again is complementary to our result \cite{RanadeAlber-2}, that a state can indeed be corrected,
if the left-hand side is strictly positive, leaving apart the case where we have equality.
\par To conclude this letter, we have used a criterion for the symmetric extendibility
(Theorem \ref{SymExt}) within the class of $\fU_2$-invariant two-qudit states in order to show that standard
two-way error-correction procedures ($\Bnd$ steps) cannot be used to improve on the already known constructive
bounds stated e.\,g. in \cite{Chau05,NikolopoulosRanadeAlber,RanadeAlber-2} by any type of one-way communication.
This was done by choosing a well-suited class of matrices for that problem, which enabled us to derive upper
bounds; these bounds coincide with the previously known bounds, thereby showing their sharpness.
In particular, our results coincide with those of Myhr \etal{} \cite{Myhr_ua} in the case of qubits ($d = 2$).
\par The author thanks \pers{Gernot Alber}, \pers{Matthias Christandl}, \pers{Norbert Lütkenhaus},
\pers{Geir Ove Myhr} and \pers{Joseph M. Renes} for helpful discussions. He was supported by a graduate-student
scholarship (Promotions\-stipendium) of the Technische Universität Darmstadt; financial support by the CASED
project is also acknowledged.

\end{document}